\documentclass{article}

\usepackage{PRIMEarxiv}
\usepackage[table,xcdraw]{xcolor}
\usepackage[utf8]{inputenc} 
\usepackage[T1]{fontenc}    
\usepackage{hyperref}       
\usepackage{url}            
\usepackage{booktabs}       
\usepackage{amsfonts}       
\usepackage{nicefrac}       
\usepackage{microtype}      
\usepackage{lipsum}
\usepackage{graphicx}
\graphicspath{{media/}}     

\begin{document}

\title{Artificial Intelligence Powered Material Search Engine}

\author{
  Divy Patel\\
  School of Technology\\
  Pandit Deendayal Energy University\\
  Gandhinagar, Gujarat, India 382007 \\
  \texttt{divypatel06122000@gmail.com} \\
   \And
 Dhairya Shah \\
  School of Technology\\
  Pandit Deendayal Energy University\\
  Gandhinagar, Gujarat, India 382007 \\
  \texttt{stariate@ee.mount-sheikh.edu} \\
   \And
 Param Modi \\
 School of Technology\\
  Pandit Deendayal Energy University\\
  Gandhinagar, Gujarat, India 382007 \\
  \texttt{stariate@ee.mount-sheikh.edu} \\
   \And
 *Mohendra Roy \\
 School of Technology\\
  Pandit Deendayal Energy University\\
  Gandhinagar, Gujarat, India 382007 \\
  \texttt{*Correspond to: mohendra.roy@ieee.org; mohendra.roy@sot.pdpu.ac.in} \\
}

\maketitle

\begin{abstract}
Many data-driven applications in material science have been made possible because of recent breakthroughs in artificial intelligence(AI). The use of AI in material engineering is becoming more viable as the number of material data such as X-Ray diffraction, various spectroscopy, and microscope data grows. In this work, we have reported a material search engine that uses the interatomic space (d value) from X-ray diffraction to provide material information. We have investigated various techniques for predicting prospective material using X-ray diffraction data. We used the Random Forest, Naive Bayes (Gaussian), and Neural Network algorithms to achieve this. These algorithms have an average accuracy of 88.50\%, 100.0\%, and 88.89\%, respectively. Finally, we combined all these techniques into an ensemble approach to make the prediction more generic. This ensemble method has a ~100\% accuracy rate. Furthermore, we are designing a graph neural network (GNN)-based architecture to improve interpretability and accuracy. Thus, we want to solve the computational and time complexity of traditional dictionary-based and metadata-based material search engines and to provide a more generic prediction.

\end{abstract}


\copyright Accepted in Materials Today: Proceedings on 4 Jan 2022. \footnote{\copyright    Materials Today Journal, Elsevier; Article reference No:  MATPR29663, DOI: \url{https://doi.org/10.1016/j.matpr.2022.01.120}}

\section{Introduction}
Recently, we have witnessed significant growth in the development of artificial intelligence (AI) and machine learning (ML) algorithms \cite{r1}. This is due to the exponential advancement in the field of accelerated compute engines and abundance in the data. This development has facilitated many applications in almost all domains of science and engineering. Material engineering is one such field, where AI can play a key role. We can explore the data-driven models to find out suitable materials for a specific application or to suggest composition in engineering a material. In this regards a material search engine may play a key role. Therefore, the development of a material search engine, with high prediction accuracy is very crucial. Many research groups are working on developing more robust and accurate search engines. However, the accuracy, time and computational complexity of such systems are yet to be explored. Moreover, the prediction of a material with unknown d-values is still a challenging issue. Most of the available material search engines are based on traditional dictionary based approach, which is well kwon for its computational cost and time complexity. Further, dictionary based approaches are poor in predicting possible characteristics of engineered materials. In this work, we have explored various AI and ML algorithms for their accuracy in predicting prospective material using the d-value of X-ray diffraction (XRD) spectroscopy. XRD is a well-known technique to characterize a crystalline material. This method can provide the insight of the crystal structure of a material by providing the information of their lattice plane spacing (in terms of d value) and thereby orientations

\textcolor{cyan}{The schematic of the proposed method is as shown in the figure 1}. Here, we have used three completely different approaches, such as bootstrap aggregation method, statistical method and biologically inspired method to explore the feasibility of the proposed method from different prospective. The detail of the methods and results are discussed below.
 \begin{figure}[ht]
\includegraphics[width=11cm]{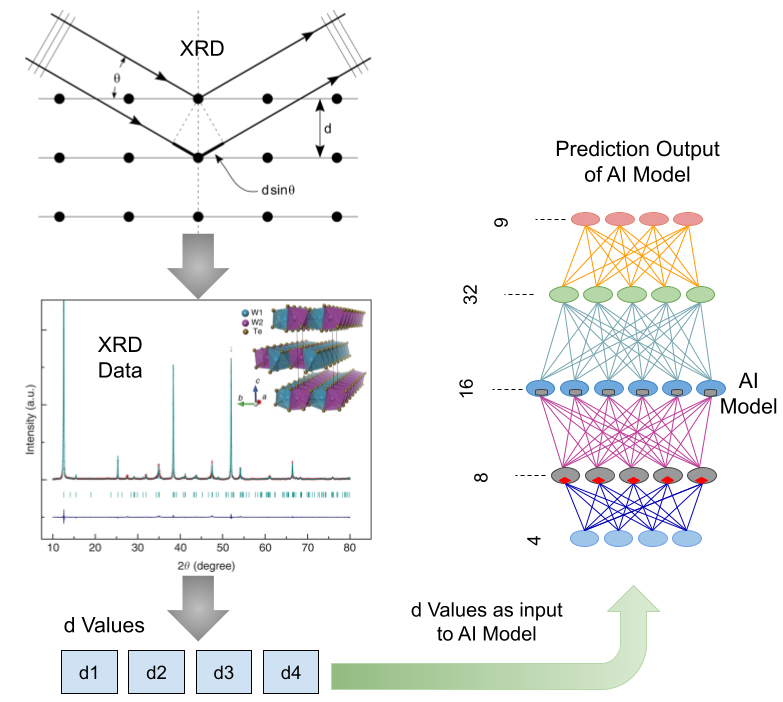}
\centering
\caption{Schematic of the proposed method of AI powered material search engine.}
\label{fig7: abc}

\end{figure}

\section{Methods}
\label{sec:headings}

\subsection{Dataset}
We have used the dataset of d-value of the material, namely Cesium Chloride, Calcium borate, Bismuth selenide, Lead, Cadmium Oxide, Selenium Oxide, Lead Oxide, Zinc Oxide and Copper Sulfide. The d-values were mostly obtained from the Inorganic Material Database \cite{r2}. The prominent four d-values were considered for each material. Thus, a dataset of 89 samples was generated.

\subsubsection{Algorithm}
We have used three different approaches, such as bootstrap aggregation method e.g. random forest, statistical method e.g. Naïve Bayes, and biologically inspired method, e.g. Neural Network. We have investigated these completely different approaches to get rid of any biasness of the underline methods of the models towards the data. Finally we have ensemble these methods to further nullify any remaining biasness. The detail of these methods are as given below.

\textbf{Random Forest Classifier:} The Random Forest Classifier is a bootstrap aggregation method \cite{r3}. It is an ensemble of decision trees \cite{r4}, which is familiar for its generic nature, as it tries to avoid overfitting by averaging the results from the trees with various hyper-parameters. The hyper-parameters for this implementation are tuned to the following: number of folds = 3, maximum branch = 18, minimum sample = 1, and sub-sample ratio = 1.0. We have evaluated this algorithm for the various number of decision trees, i.e., for 1, 5, 8, and 21.

\textbf{Naive Bayes (gaussian):} Naïve Bayes (NB) is statistical method and a probabilistic classifier \cite{r5}. The probability density of a given class $C_{k}$ or some observation value $v$ is given by $p(x={v}|{C_{k}})$. This can be evaluated by mean ($\mu _{k}$) and variance ($\sigma _{k}^2$) of the values associated with the class $C_{k}$ which is given by $p(x={v}|{C_{k}})=\frac{1}{\sqrt{2\pi\sigma {k}^{2} }}e^{-\frac{(\upsilon - \mu )^{2}}{2\sigma {k}^{2}}}$. In this work, there were no priors set for the dataset. For multiple classes, we take precision and recall of each class and take its average to calculate the F-score

\textbf{Neural Network:} The Neural network is a biologically inspired network of artificial \cite{r6}. The network learns the features from the input by adjusting the weights in the networks. The schematic of the custom network is as shown in \textcolor{cyan}{Figure 2}. In our work, we used the fully connected network with three hidden layers. The input layer is consist of four neurons (for each d-values) and the output layer has nine neurons (for each class of materials). The hidden layers contain 8, 16 and 32 neurons respectively.

\begin{figure}[ht]
\includegraphics[width=8cm]{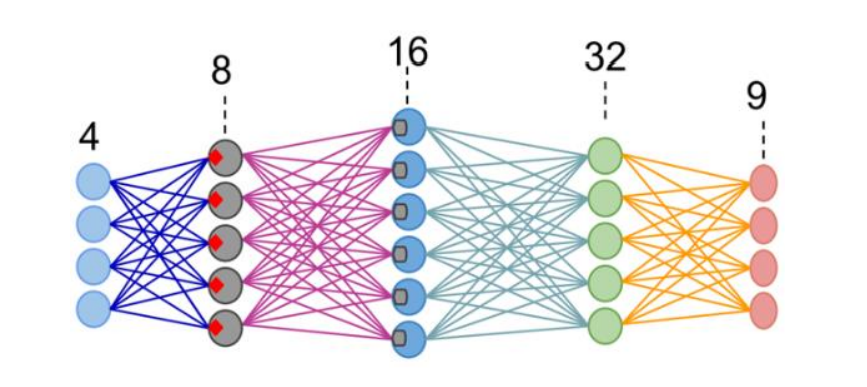}
\centering
\caption{Schematic of the proposed neural network.}
\label{fig2: Schematic of the proposed neural network.}

\end{figure}

\textbf{Ensemble Model:} In the ensemble model \cite{r7}, we have combined the probability score from the Random Forest Classifier, Naive Bayes, and Neural Network. The highest average probability among the classes is considered as the predicted class.

\textbf{F-Score:} We have used the standard F-score as the evaluation metric for the performance of the models. To get a clear picture, we have used the F1 score along with accuracies. F1 score is based on the precision and recall of the model. The F1 score is calculated using the following equation.

\begin{center}
   $F_{1} = 2\frac{precision*recall}{precision+recall}=\frac{TP}{TP+\frac{1}{2}(FP+FN)}$
 
\end{center}

Here,
TP = True Positives - These are the values which are correctly predicted true by the model 
\\
FP = False Positives - These are the values which are incorrectly predicted true by the model
\\
 FN = False Negatives - These are the values which are incorrectly predicted false by the model

\section{Result and discussion}

The comparison of the results for all the three algorithms and their ensembles are as shown in table 1. From the comparison, it is clear that the Naive Bayes classifier and the ensemble method is performing better. The F1 score is indicating that the models are performing well in balance for all the classes in NB and Ensemble. However, Random Forest and Neural Network still have rooms for improvement. Since we are using the average probability of classes in the ensemble method, therefore, we may consider the model as more generic, with higher accuracy.

\begin{table}[h!]
\centering
\caption{Comparison of results from various algorithms in predicting the material from their d-value.}
\label{tab:my-table}
\begin{tabular}{cccccc}
\rowcolor[HTML]{9B9B9B} 
\textbf{Model}       & \textbf{\begin{tabular}[c]{@{}c@{}}Train Time \\ (in seconds)\end{tabular}} & \textbf{\begin{tabular}[c]{@{}c@{}}Train accuracy \\ (\%)\end{tabular}} & \textbf{\begin{tabular}[c]{@{}c@{}}Test accuracy \\ (\%)\end{tabular}} & \textbf{Train F1 Score} & \textbf{Test F1 Score} \\
Random Forest        & 0.021                                                                       & 89                                                                      & 88.50                                                                  & 0.87                    & 0.85                   \\
\rowcolor[HTML]{C0C0C0} 
Naive Bayes Gaussian & 0.0032                                                                      & 100                                                                     & 100                                                                    & 1                       & 1                      \\
Neural Network       & 37.6685                                                                     & 88.57                                                                   & 88.89                                                                  & 0.84                    & 0.84                   \\
\rowcolor[HTML]{C0C0C0} 
Ensemble             & 37.6928                                                                     & 100                                                                     & 100                                                                    & 1                       & 1                     
\end{tabular}
\end{table}

\section{Conclusion}

In conclusion, we have explored various machine learning and AI models. This study indicates that we can exploit the machine learning/ AI models for designing more generic material search engines. From this study we found that ensemble of bootstrapping method like Random forest, Statistical method such as Naïve Bays and biologically inspired artificial Neural Network provides more generic and accurate prediction of materials from their d-values with about 100\% accuracy and highest F-score. In future, we are going to study the computation and time complexity of these models. Currently, we are developing an explainable material search engine using Graph Neural Network. This may provide more insight while proposing a synthetic material. Further, the limitation of the data may be handle by incorporating data augmentation techniques or by incorporating more number of materials in future.

\section{Acknowledgements}
The authors acknowledge the seed grant provided by ORSP, PDEU, for the seed grant project No: ORSP/ R\&D/ PDPU/2019/MR/RO051.

\bibliographystyle{unsrt}

\end{document}